\documentclass[twoside,showpacs,superscriptaddress,twocolumn,floatfix,a4paper,aps,prl]{revtex4}

\usepackage{color}
\usepackage{graphicx}
\usepackage[utf8x]{inputenc}
\usepackage[T1]{fontenc}
\usepackage{siunitx}
\usepackage{amstext}
\usepackage{amsmath,amssymb}
\usepackage{textgreek}
\usepackage{hyperref}
\usepackage{dcolumn}

\newcommand{\etal}{\textit{et al.~}}

\newcommand{\one}{\leavevmode\hbox{\small1\normalsize\kern-.33em1}}

\newcommand\kb[1]{#1}
\newcommand\kbq[1]{#1}

\begin{document}

\title{Experimental implementation of optimal linear-optical controlled-unitary gates}

\author{Karel Lemr}
\email{k.lemr@upol.cz}
\affiliation{RCPTM, Joint Laboratory of Optics of Palacký University and Institute of Physics of Academy of Sciences of the Czech Republic, 17. listopadu 12, 771 46 Olomouc, Czech Republic}

\author{Karol Bartkiewicz} \email{bartkiewicz@jointlab.upol.cz}
\affiliation{Faculty of Physics, Adam Mickiewicz University,
PL-61-614 Pozna\'n, Poland}
\affiliation{RCPTM, Joint Laboratory of Optics of Palacký University and Institute of Physics of Academy of Sciences of the Czech Republic, 17. listopadu 12, 771 46 Olomouc, Czech Republic}

\author{Antonín Černoch} \email{acernoch@fzu.cz}
\affiliation{Institute of Physics of Academy of Sciences of the Czech Republic, Joint Laboratory of Optics of PU and IP AS CR, 
   17. listopadu 50A, 772 07 Olomouc, Czech Republic}

\author{Miloslav Dušek}
\affiliation{Department of Optics, Faculty of Science, Palacký University, 
   17. listopadu 12, cz-77146 Olomouc, Czech Republic}

\author{Jan Soubusta}
\affiliation{Institute of Physics of Academy of Sciences of the Czech Republic, Joint Laboratory of Optics of PU and IP AS CR, 
   17. listopadu 50A, 772 07 Olomouc, Czech Republic}

\date{\today}

\begin{abstract}
We show that \kb{it is possible to} reduce the number of two-qubit gates needed for \kb{the} construction of an arbitrary controlled-unitary transformation by up to two times \kb{using a tunable controlled-phase gate}. On the platform of linear optics, where two-qubit gates can only be achieved probabilistically, our method significantly reduces \kb{the amount of components and increases success probability of a two-qubit gate. The} experimental \kb{implementation} of our technique \kb{presented in this paper} for \kb{a} controlled single-qubit unitary gate \kb{demonstrates that} only  one tunable controlled-phase gate \kb{is needed} instead of two standard controlled-NOT gates. Thus\kb{,} not only \kb{do} we increase success probability by about one order of magnitude (with the same resources), but also avoid the need for \kb{conducting} quantum non-demolition measurement otherwise required to join two probabilistic gates. Subsequently, we generalize our method to \kb{a} higher order\kb{,} showing that $n$-times controlled gates can be optimized \kb{by} replacing blocks of controlled-NOT gates with tunable controlled-phase gates.
\end{abstract}


\maketitle

Quantum computing is a promising \kb{direction in} information processing \cite{Nielsen_QCQI,Zeilinger_QIP}. Similarly to classical computing, quantum circuits are composed of various elementary gates.  In 1989\kb{,} Deutsch proved \kb{the} existence of a universal three-qubit gate \cite{Deutsch89three}\kb{. Later,} DiVincenzo showed that Deutsch's gate can be implemented by a sequence of two- and single-qubit gates \cite{DiVincenzo95two}. Meanwhile\kb{,} Barenco discovered a class of two-qubit gates sufficient for building any quantum circuit \cite{Barenco94universal}. \kb{A} practical set of universal gates was defined 
\kb{later} \cite{Barenco95universal}. This set of gates includes several single-qubit gates and only one two-qubit gate --- the controlled-NOT (CNOT) gate.

Although the method presented by Barenco \etal \cite{Barenco95universal} shows how to construct any quantum circuit, it does not take into account various \kb{optimization} procedures \cite{Bullock03optim,Zhang03optim,Shende04optim,Vartiainen04optim,Mottonen04optim,Vatan04optim,Vidal04optim,Bergholm05optim,Shende06optim,Plesch11optim}.  \kb{Optimization} is crucial \kb{in linear optics}, \kb{where the} CNOT gate can only be implemented probabilistically \cite{Hofmann02cpg,Langford05cpg,Kiesel05cpg,Bartkowiak10cpg}, meaning that  every repetition reduces the success probability of the entire scheme.

In 2009, Lanyon \etal demonstrated a considerable reduction in \kb{the} number of CNOT gates \kb{necessary for} circuit construction by introducing additional ancillary modes \cite{Lanyon09cpg}. \kb{They}  have constructed a Toffoli gate (controlled-controlled-NOT gate) with only two CNOT gates and have also \kb{designed} a generalized controlled-phase gate, but not an optimal one. Mičuda \etal presented a method further reducing resources needed for \kb{the} implementation of a Toffoli gate to only one CNOT gate \cite{Micuda13toffoli}. This reduction is achieved by combining polarization and spatial encoding to encode \kb{a} two-qubit state into one single photon. However, \kb{the preparation of a} specialized control two-qubit state is problematic. \kb{It is possible to} use quantum routers (or quantum state fusion) \cite{Lemr13router,Lemr13cpg_router,Vitelli2013fusion}, but \kb{this would mean using} additional CNOT gates\kb{, which would} cancel the achieved reduction.

So far only optimizations involving standard CNOT or controlled-sign (c-sign) gates were considered. 
In 2010, Kieling \etal proposed an optimal (without auxiliary photons) linear-optical implementation 
of a tunable c-phase gate that imposes a given tunable phase shift $\varphi$  \cite{Kieling10cpg}
\begin{eqnarray}
\label{eq:cpg_def}
|kl\rangle & \rightarrow & \mathrm{e}^{\mathrm{i}\varphi\delta_{k1}\delta_{l1}} |kl\rangle, 
\end{eqnarray}
where $k$ and $l$ take values of logical qubit states $0$ or $1$ and $\delta$ is the Kronecker's delta. In 2011, \kb{the} first optimal tunable c-phase gate was experimentally \kb{demonstrated} \cite{Lemr11cpg}. \kb{The} experiment also allowed to verify and explain \kb{the} optimal success probability of \kb{the} gate as \kb{a} function of the phase shift $\varphi$ \cite{Kieling10cpg,Lemr12cpg_prob}.

In this \kb{Letter} we show that using a tunable c-phase gate instead of a CNOT gate \kb{makes it possible}  to (i) reduce \kb{the} complexity of various quantum circuits and (ii) increase \kb{the} success probability of these circuits \kb{in} linear optics. \kb{The support for our idea comes from an experimental implementation of the proposed scheme}. 

{\em Arbitrary single-qubit controlled-unitary transformation} -- It has been shown by Barenco \etal \cite{Barenco95universal} that two controlled-sign gates are needed to implement an arbitrary controlled-unitary operation acting on a signal qubit and controlled by a control qubit. In special cases, \kb{} one controlled-sign gate is sufficient, but at the expense of restricting the class of implemented operations. Considering the probabilistic nature of controlled-sign gates on the platform of linear optics, it is crucial to limit their repetition as much as possible. We show  that only one single tunable controlled-phase gate is needed for \kb{the} construction of a universal single-qubit controlled-unitary operation. Note\kb{,} that \kb{the success probability of} two consecutive controlled-sign gates would \kb{be} $1/81$ (using linear optics only and no photon ancillae), the \kb{minimum success probability of a} tunable controlled-phase gate \kb{is} $1/11$ (0.14 on average). Moreover\kb{,} by reducing the number of gates from two to one, we also avoid the need for intermediary non-demolition presence detection otherwise required to join two probabilistic gates \cite{Kok02qnd,Bula13qnd}. 

\begin{figure}
\includegraphics[scale=1]{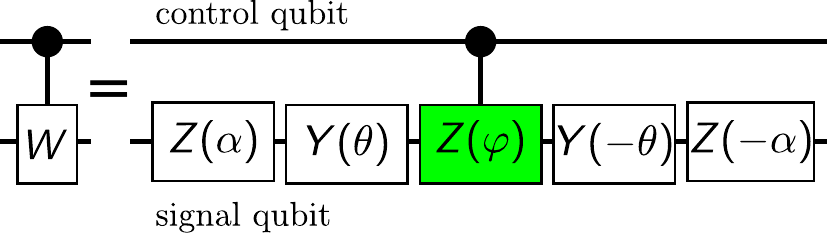}
\caption{\label{fig:universal} (color online) Quantum computation circuit implementing an arbitrary single-qubit controlled-unitary operation $W$ [see Eq. (\ref{eq:W})] by means of one tunable controlled-phase gate and several unconditional single-qubit operations.}
\end{figure}

Let us consider the scheme depicted in Fig. \ref{fig:universal}. While the upper (control) qubit undergoes only the controlled-phase operation, the lower (signal) qubit is subjected to a set of unconditional single-qubit gates before and after it enters the controlled-phase gate. These unconditional single-qubit gates can be implemented deterministically on the platform of linear optics.

The initial set of single-qubit operations consists of one rotation in $z$ direction
\begin{equation}
\label{eq:Rz}
Z(\alpha) =
\begin{pmatrix}
\mathrm{e}^{-\mathrm{i}\alpha/2} & 0 \\
0 & \mathrm{e}^{\mathrm{i}\alpha/2}
\end{pmatrix}
\end{equation}
followed by another rotation in $y$ direction
\begin{equation}
\label{eq:Ry}
Y(\theta) =
\begin{pmatrix}
\cos\frac{\theta}{2} & \sin\frac{\theta}{2} \\
-\sin\frac{\theta}{2} & \cos\frac{\theta}{2} \\
\end{pmatrix}.
\end{equation}
Similarly, the single-qubit rotations inserted behind the controlled-phase gate are $Y(-\theta)$ and $Z(-\alpha)$. When the control qubit is $|0\rangle$, the controlled-phase gate does not impose any phase \kb{shifts} and all unconditional single-qubit rotations cancel each other
$$
Z(-\alpha)Y(-\theta)Y(\theta)Z(\alpha) = \one.
$$

On the other hand, if the control qubit is $|1\rangle$, the controlled-phase gate introduces an additional rotation in $z$ direction $Z(\varphi)$ [see Eq. (\ref{eq:Rz})]. The overall operation imposed on the signal qubit now reads
\begin{equation}
\label{eq:W}
W = Z(-\alpha)Y(-\theta)Z(\varphi)Y(\theta)Z(\alpha).
\end{equation}

To demonstrate the universality of the above mentioned gate, let us consider the following: Any single-qubit unitary transformation can be described as a rotation along some axis on the Bloch sphere which corresponds to an operator
\begin{equation}
R_{\psi}(\varphi) = \mathrm{e}^{\mathrm{i}\varphi/2}|\psi\rangle\langle \psi| + \mathrm{e}^{-\mathrm{i}\varphi/2}|\psi^\perp\rangle\langle \psi^\perp|,
\end{equation}
where $\varphi$ denotes the rotation angle and $|\psi\rangle$ is the state that geometrically corresponds to the rotation axis on the Bloch sphere ($|\psi^\perp\rangle$ is orthogonal to $|\psi\rangle$). For rotation along the $z$ direction we have $|\psi\rangle = |0\rangle$ which inserted to (\ref{eq:W}) yields
\begin{equation}
W = Z(-\alpha)Y(-\theta)\left(\mathrm{e}^{\mathrm{i}\varphi/2}|0\rangle\langle 0| + \mathrm{e}^{-\mathrm{i}\varphi/2}|1\rangle\langle 1|\right)Y(\theta)Z(\alpha).
\end{equation}
Using prescriptions (\ref{eq:Rz}) and (\ref{eq:Ry}) we can easily verify that
\begin{equation}
|\psi\rangle = Z(-\alpha)Y(-\theta)|0\rangle = \mathrm{e}^{\mathrm{i}\frac{\alpha}{2}}\cos\frac{\theta}{2}|0\rangle + \mathrm{e}^{-\mathrm{i}\frac{\alpha}{2}}\sin\frac{\theta}{2}|1\rangle,
\end{equation}
and thus show that any arbitrary pure qubit state is accessible \kb{if} suitable values of $\alpha$ and $\theta$
\kb{are set}. Unitary transformations maintain orthogonality so that $|1\rangle\rightarrow Z(-\alpha)Y(-\theta)|1\rangle = |\psi^\perp\rangle$. The two pairs of unconditional single-qubit rotations \kb{before and after} the c-phase gate \kb{permit for} any rotation axis. \kbq{Tunning the phase of} the c-phase gate \kbq{permits for setting} any rotation angle $\varphi$.

Any single-qubit unitary operation can be decomposed in the form of \cite{Barenco95universal}
\begin{eqnarray}
\label{eq:unitary_docomposed}
U = Z(\gamma)Y(\omega)Z(\delta)
\end{eqnarray}
parametrized by three real numbers. \kb{An} explicit decomposition of \kb{the} transformation  matrix \kb{can be found} in the Supplementary material~\cite{supp}. Note that the matrix $W$ in (\ref{eq:W}) is also parametrized by three real numbers. Optimality of our method is guaranteed by the the fact that we only use one probabilistic gate which is optimal for any given phase shift required by the transformation.


\begin{figure}
\includegraphics[scale=1]{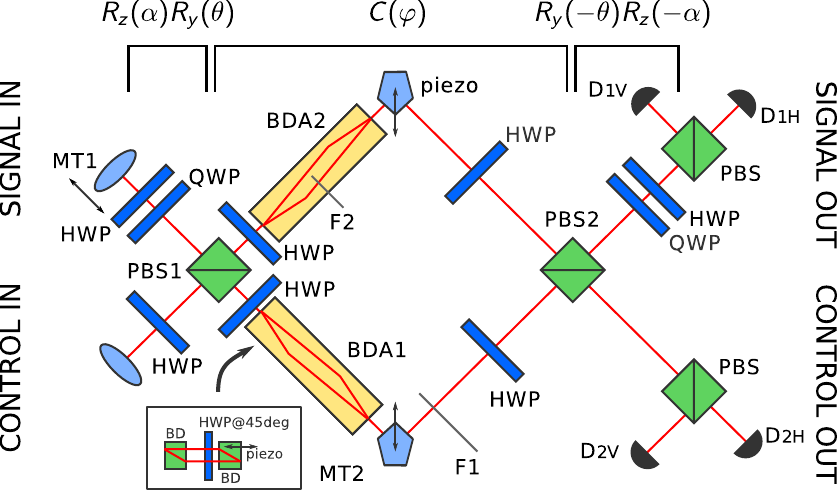}
\caption{\label{fig:setup} (color online) \kb{Schematic drawing} of the experimental setup. The components are \kb{labeled} as follows: MT -- motorized translation, HWP -- half-wave plate, QWP -- quarter-wave plate, PBS -- polarizing beam splitter, BDA -- beam divider assembly, BD -- beam divider, D -- detector.}
\end{figure}


{\em Experimental implementation} -- We have constructed an experimental setup as depicted in Fig. \ref{fig:setup}. It consists of a tunable c-phase gate \kbq{placed between single-qubit gates} in the signal mode \kbq{that implement the} required unconditional rotations $Z$ and $Y$. In our experiment we encode qubits into polarization states of individual photons ($|0\rangle$ corresponds to horizontal polarization $|H\rangle$, $|1\rangle$ to vertical polarization $|V\rangle$). Unconditional single-qubit rotations $Z$ and $Y$ are implemented by sets of one half- and one quarter-wave plates.
The control state preparation is achieved by one half-wave plate \kb{} in control mode since only logical states $|0\rangle$ and $|1\rangle$ are required. Photons were generated using Type I spontaneous parametric down-conversion in a LiIO$_3$ crystal pumped by 200\,mW cw Kr$^+$ laser beam. By following the procedure described in Ref. \citep{Lemr11cpg}, we have adjusted the tunable c-phase gate to a given phase shift $\varphi$.

\begin{figure}
\includegraphics[scale=0.85]{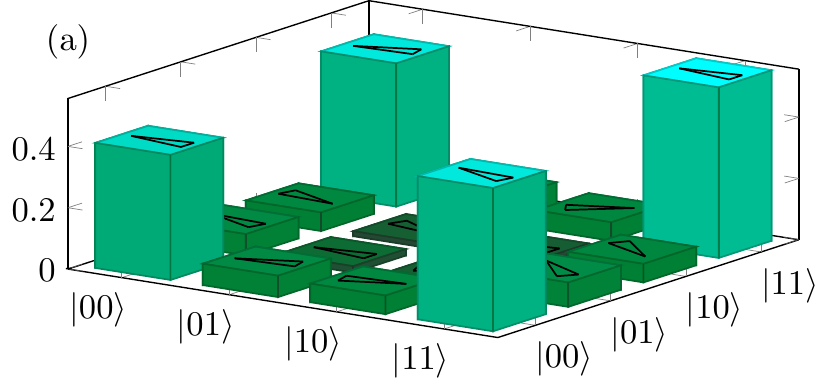}
\includegraphics[scale=0.85]{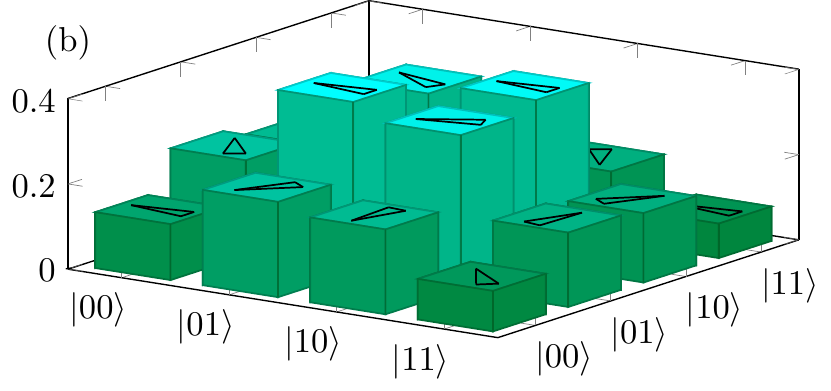}
\caption{\label{fig:res_cuD3pi4} (color online) Estimated process matrices for $\varphi=\frac{3\pi}{4}$, $\theta=\frac{\pi}{2}$ and $\alpha=0$ (a) with control qubit $|0\rangle$ and (b) with control qubit $|1\rangle$. 
Moduli of matrix elements are visualized by bar heights and their phase by arrow directions.}
\end{figure}

We have tested our device on six combinations of tunable c-phase gate phase shifts $\varphi$ and  single qubit rotations $Z(\alpha)$ and $Y(\theta)$ (see Tab. \ref{tab:res_cunit}). In all these six cases, \kb{we have performed} complete process tomography of the signal mode for \kb{the} control qubit set to state $|0\rangle$ and then also to $|1\rangle$ \cite{Jamiolkowski1972iso,Choi1975iso,Jezek2003estimace,Paris2004estimace}. The estimated Choi matrices were compared to theoretical predictions \kb{permitting} to calculate their fidelities $\mathcal{F}$ and purities $\mathcal{P}$. We adopt the following \kb{labeling}: $\mathcal{F}_\mathrm{off}$ and $\mathcal{P}_\mathrm{off}$ stand for fidelity and purity observed with control qubit set to $|0\rangle$, while $\mathcal{F}_\mathrm{on}$ and $\mathcal{P}_\mathrm{on}$ denote the same parameters for control \kb{qubits} in the state $|1\rangle$. We have also determined the resulting success probabilities by comparing the coincidence rate observed after adjusting the gates with the coincidence rate behind the same setup, but with all filters removed and polarizations set so that no single or two-photon interference takes place. Thus\kb{,} we obtain the experimental success probability $p_\mathrm{succ}$ corrected for ``technological losses'' (e.g. components back-reflections or coupling losses). \kb{The r}esults of our experiment are summarized in Tab.~\ref{tab:res_cunit} and one selected case is also depicted in Fig.~\ref{fig:res_cuD3pi4}. Estimated fidelities and purities are typically about 90\% which indicates good agreement with theoretical predictions.


\begin{table*}

\caption{\label{tab:res_cunit}Experimental results for various settings of \kb{the} c-phase gate parameter $\varphi$ together with single qubit rotations $\alpha$ and $\theta$. The corresponding values of standard decomposition parameters $\omega$, $\gamma$ and $\delta$ are also calculated. $\mathcal{F}_\mathrm{off}$ and $\mathcal{P}_\mathrm{off}$ denote estimated process fidelity and purity of the transformation with \kb{the} control qubit set to $|0\rangle$, while $\mathcal{F}_\mathrm{on}$ and $\mathcal{P}_\mathrm{on}$ denote the same characteristics with \kb{the} control qubit set to $|1\rangle$.  $p_\mathrm{succ}$ and $p_\mathrm{succTH}$ stand for experimental and theoretical success probability.}

\begin{ruledtabular}
\begin{tabular}{rrrrrrcccccc}
$\varphi$ & $\theta$ & $\alpha$ & $\omega$ & $\gamma$ & $\delta$ & $\mathcal{F}_\mathrm{off}$ & $\mathcal{P}_\mathrm{off}$ & $\mathcal{F}_\mathrm{on}$ & $\mathcal{P}_\mathrm{on}$ & $p_\mathrm{succ}$ & $p_\mathrm{succTH}$\\\hline

0 & 0 & 0 & 0 & 0 & 0 & $0.964\pm 0.001$ & $0.940\pm 0.001$ & $0.971\pm 0.001$ & $0.957\pm 0.001$ & $0.822\pm 0.059$ & $1.000$ \\ 

${\pi}/{8}$ & ${\pi}/{2}$ & $0$ & ${\pi}/{8}$ & ${\pi}/{2}$ & $-{\pi}/{2}$ & $0.973\pm 0.001$ & $0.981\pm 0.003$ & $0.918\pm 0.002$ & $0.912\pm 0.005$ & $0.180\pm 0.008$ & $0.210$ \\ 

${\pi}/{4}$ & $0$ & ${\pi}/{2}$ & $0$ & ${\pi}/{8}$ & ${\pi}/{8}$ & $0.928\pm 0.003$ & $0.921\pm 0.005$ & $0.878\pm 0.004$ & $0.863\pm 0.006$ & $0.105\pm 0.017$ & $0.133$ \\ 

${\pi}/{2}$ & ${\pi}/{2}$ & ${\pi}/{2}$ & ${\pi}/{2}$ & $0$ & $0$ & $0.900\pm 0.004$ & $0.909\pm 0.006$ & $0.909\pm 0.004$ & $0.913\pm 0.007$ & $0.086\pm 0.005$ & $0.090$ \\ 

${3\pi}/{4}$ & ${\pi}/{2}$ & $0$ & ${3\pi}/{4}$ & ${\pi}/{2}$ & $-{\pi}/{2}$ & $0.937\pm 0.003$ & $0.950\pm 0.006$ & $0.889\pm 0.004$ & $0.909\pm 0.007$ & $0.079\pm 0.006$ & $0.088$ \\ 

$\pi$ & $0$ & ${\pi}/{2}$ & $0$ & ${\pi}/{2}$ & ${\pi}/{2}$ & $0.934\pm 0.004$ & $0.936\pm 0.006$ & $0.771\pm 0.005$ & $0.833\pm 0.007$ & $0.108\pm 0.016$ & $0.111$ \\ 
\end{tabular}
\end{ruledtabular}

\end{table*}


{\em $n$-times controlled single qubit unitary transformations} -- As in case of a CNOT gate, a Toffoli gate (CCNOT) can be used to implement a controlled unitary gate, but with two control qubits. The CCNOT operation can be implemented using only CNOT gates. We can also build a $2$-times controlled gate by replacing the CNOT gates acting on the target qubit with \kb{} c-phase gates and single-qubit rotations. This approach \kb{ensures}  efficiency \kb{higher} than \kb{in case of a} circuit using only CNOTs, where CNOT gates modifying the target qubit are each replaced with a single-qubit rotation sandwiched \kb{between} two CNOT gates. This means the latter approach \kbq{is equivalent to adding} 3 two-qubit gates \kbq{to} the circuit \kbq{proposed} in Lemma~6.1 in Ref.~\cite{Barenco95universal}.

The situation is analogous for any $n$-times controlled unitary gate. Designing an arbitrary $n$-times controlled gate is usually considered in the context of Toffoli gates. However, constructing such circuits directly with Toffoli gates has been proven \kb{} inefficient as it needs an order of $n^2$ two-qubit gates \cite{Nielsen_QCQI}. It has been demonstrated that\kb{,} by extending \kb{the} Hilbert space of the target information carrier (see Ref.~\cite{Lanyon09cpg})\kb{,} one can implement an arbitrary $n$-times controlled gate by using $2n-1$ standard two-qubit gates performing controlled-Pauli operations ($R_n(\pi)$ for $n=x,y,z$). This is considered to be the most effective \kb{currently known} solution. 

\begin{figure}
\includegraphics[width=8cm]{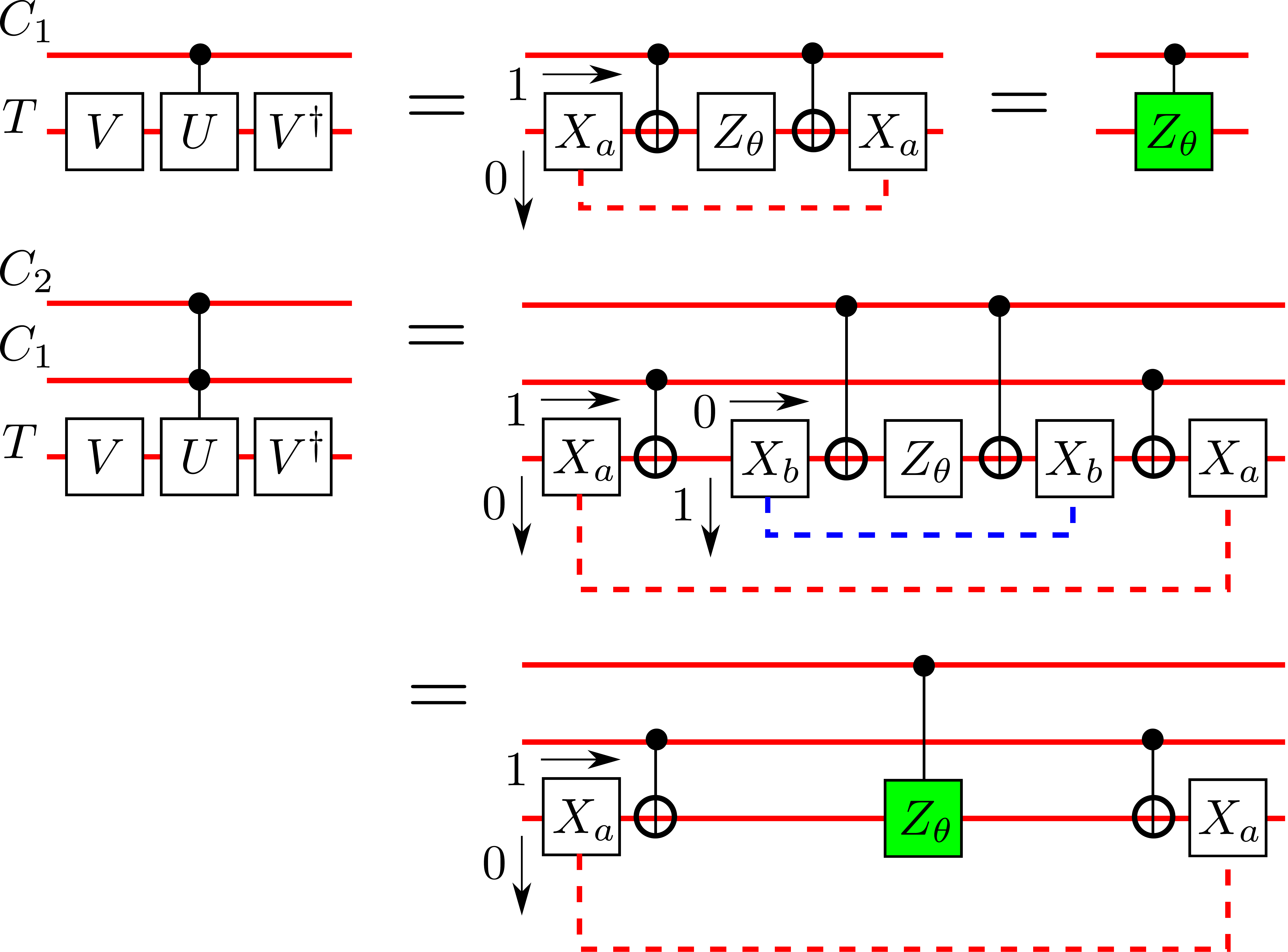}
\caption{\label{fig:L09nU}(color online) 2-times and 3-times c-phase [$Z(\theta)$] gates decomposed with a method of Lanyon~{\it et al.}~\cite{Lanyon09cpg} based on extending the Hilbert space of the target qubit. The control lines are denoted as $C_n$ for $n=1,2,3$ and $T$ denotes the target qubit line. For the 2(3)-times controlled gate the dimension of the target qubit is extended by 1(2). The additional dimensions are visualized \kb{as} additional dashed channels. For technical details on operation of $X_a$ and $X_b$ gates see Ref.~\cite{Lanyon09cpg}. By iterating the depicted procedure for an $n$-times controlled gate we conclude that $2(n-1)$ CNOTs and a single c-phase are needed for its implementation. An arbitrary controlled-$U$ gate, where $U = V^\dagger Z_{\theta}V$ can be \kb{easily} implemented \kb{using} the same circuits by applying $V$ and $V^\dagger$ single-qubit operations.}
\end{figure}

\kb{In} linear optics we can increase the efficiency of an $n$-times controlled unitary gate proposed by Lanyon \etal~\cite{Lanyon09cpg} by replacing two standard controlled-Pauli gates with a controlled-unitary operation performed with higher \kb{probability of success}. In the original scheme, the authors of Ref.~~\cite{Lanyon09cpg} show that any $n$-qubit controlled gate can be implemented using only $2n$ CNOT gates and single-qubit operations. Similarly, we show in Fig.~\ref{fig:L09nU} any $n$-times controlled-$U$ gate can be reduced to $2(n-1)$ CNOT gates and one single-qubit controlled-phase gate. This is possible by using $(n+1)$-level system as the bottom-most circuit line. This reduces the number of the required two-qubit gates by one\kb{,} and has the additional merit \kb{in having the} single qubit controlled-$U$ gate work\kb{} with \kb{a} higher \kb{average} success probability than a CNOT gate. 

This optimization can be applied in \kb{a} linear-optical implementation of the $2$-times controlled unitary gate from Ref.~\cite{Lanyon09cpg}. The improvement is apparent \kb{if} we consider replacing the product of CNOT, CZ and R gates by a product of two controlled-unitary gates and single-qubit operations. The product of two controlled-unitary gates works on average with \kb{a} higher success rate than a product of CZ and CNOT gates. Hence, by using an optimal implementation of \kb{a} c-phase gate\kb{,} we can increase the success rate of the circuit by approximately one order of magnitude.

The \kb{currently known} most efficient implementation of \kb{the} Toffoli gate\kb{} in terms of the number of two-qubit gates\kb{} was presented by Mičuda \etal in Ref.~\cite{Micuda13toffoli}. This approach \kb{requires} only one CZ gate. However, \kb{it is possible to} obtain \kb{an even} more efficient circuit for the controlled-$U$ if we replace the CZ with \kb{a} c-phase gate instead of the product of two CNOTs and single-qubit rotations. 

Any $n$-times controlled operation can be constructed from Toffoli gates and a single-qubit controlled unitary gate regardless of the auxiliary resources. Therefore, we \kb{infer} that \kb{our} reasoning and the resulting improvement is valid for all implementations of the CCNOT.


{\em Conclusions} --- We have presented a method for \kb{the} optimization of quantum circuits based on \kb{a} tunable c-phase gate. This method decreases the number of controlled operations needed in circuit design and \kb{} significantly increases the success probability on physical platforms\kb{} where controlled gates can only be implemented probabilistically (e.g. linear optics). Moreover\kb{,} the reduction in number of gates makes circuits less complex and thus\kb{more} experimentally \kb{} accessible. We have demonstrated \kb{the} experimental feasibility of our approach on \kb{on the basis of one} experimental case\kb{, namely} the single-qubit controlled-unitary operation.


{\em Acknowledgements} --- The authors thank Jára Cimrman for \kb{his} helpful suggestions. K.~L. acknowledges support by Czech Science Foundation (Grant No. 13-31000P). A.~Č. acknowledges support by Czech Science Foundation (Grant No. P205/12/0382). K.~B. acknowledges support by the Foundation for Polish Science and
the Polish National Science Centre under grant No.
DEC-2013/11/D/ST2/02638, the
Operational Program Research and Development for Innovations
European Regional Development Fund (Project No.
CZ.1.05/2.1.00/03.0058) and Operational Program Education for
Competitiveness European Social Fund (Project No.
CZ.1.07/2.3.00/20.0017) of the Ministry of Education, Youth and
Sports of the Czech Republic. M.~D. acknowledges support from the Palacky University (IGA-PrF-2014008).


\section*{Supplementary material}
\subsection*{Decomposition of single-qubit controlled-unitary operation}

We have demonstrated in the text, that the set of gates depicted in Fig. 1 (in the main 
text) allows to implement any controlled-unitary transformation in the signal qubit mode.
The overall operation of the gates on the signal qubit can be described by the matrix 
\begin{eqnarray}
\label{eq:W_ZYZYZ}
  W = Z(-\alpha) Y(-\theta) Z(\varphi) Y(\theta) Z(\alpha).
\end{eqnarray}
It is however customary to decompose the desired unitary transformations in the form of
\begin{eqnarray}
\label{eq:U_ZYZ}
  U = Z(\gamma) Y(\omega) Z(\delta).
\end{eqnarray}
In this section, we find analytical formulas allowing to map the set of parameters 
$\lbrace\alpha,\theta,\varphi\rbrace$, used in the matrix $W$ describing our scheme, 
to the set of decomposition parameters $\lbrace\gamma,\omega,\delta\rbrace$. 
Let us now express both the transformations $U$ and $W$ in matrix forms using 
definitions introduced in the main text. 

Firstly, we write the form of matrix $W$ in a compact form as follows 
\begin{eqnarray}
\label{eq:Wmatrix}
W & = & 
\begin{pmatrix}
\chi & \xi \\
-\xi^* & \chi^*
\end{pmatrix},
\end{eqnarray}
where we introduced two auxiliary parameters
\begin{equation}
\label{eq:chi}
 \chi (\theta,\varphi) 
 = \mathrm{e}^{-i\varphi/2}\cos^2\frac{\theta}{2}
 + \mathrm{e}^{ i\varphi/2}\sin^2\frac{\theta}{2}
\end{equation}
and
\begin{equation}
\label{eq:xi}
 \xi (\alpha,\theta,\varphi) 
 = \mathrm{e}^{i(\alpha-\frac{\pi}{2})}\sin\theta\sin\frac{\varphi}{2}.
\end{equation}

Secondly, we derive the matrix form of the decomposed unitary transformation 
\begin{eqnarray}
\label{eq:unitary_docomposed}
U = 
\begin{pmatrix}
  \mathrm{e}^{-i\frac{ \gamma+\delta}{2}}\cos\frac{\omega}{2} & 
  \mathrm{e}^{ i\frac{-\gamma+\delta}{2}}\sin\frac{\omega}{2} \\
 -\mathrm{e}^{ i\frac{ \gamma-\delta}{2}}\sin\frac{\omega}{2} & 
  \mathrm{e}^{ i\frac{ \gamma+\delta}{2}}\cos\frac{\omega}{2}
\end{pmatrix}.
\end{eqnarray}
Our goal is now to find a unique relation between both sets of three parameters,
that guarantee equivalence $U(\gamma,\omega,\delta) = W(\alpha,\theta,\varphi)$.

We start by comparing the amplitudes of the individual terms of the matrices. 
One can straightforwardly identify
\begin{eqnarray}
\label{eq:univ_amplitude1}
 \cos\frac{\omega}{2} & = & |\chi|
\end{eqnarray}
and simultaneously
\begin{eqnarray}
\label{eq:univ_amplitude2}
 \sin\frac{\omega}{2} & = & |\xi|.
\end{eqnarray}
Since the relation $|\chi|^2+|\xi|^2=1$ holds disregarding the values of $\varphi$ and 
$\theta$, one can always fulfil both conditions (\ref{eq:univ_amplitude1}) and 
(\ref{eq:univ_amplitude2}) simultaneously. These equations allow to calculate parameter 
$\omega$ 

Analysing the phases of the diagonal and offdiagonal terms directly reveals that
\begin{equation}
\label{eq:univ_phase1}
  {\delta+\gamma} = -2\sphericalangle (\chi),
\end{equation}
respectively
\begin{equation}
\label{eq:univ_phase2}
  \delta-\gamma = 2\alpha-\pi.
\end{equation}
This set of two equations can be easily solved obtaining both $\gamma$ and $\delta$ as 
functions of $\alpha$ and $\sphericalangle (\chi)$, that is the polar angle of the 
complex number $\chi$.

\begin{figure*}
\includegraphics[scale=0.88]{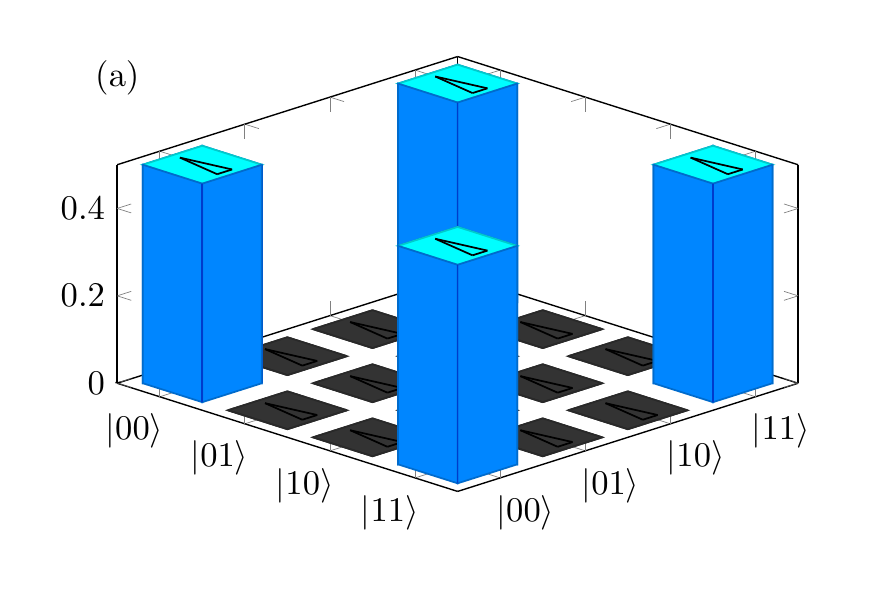} 
\includegraphics[scale=0.88]{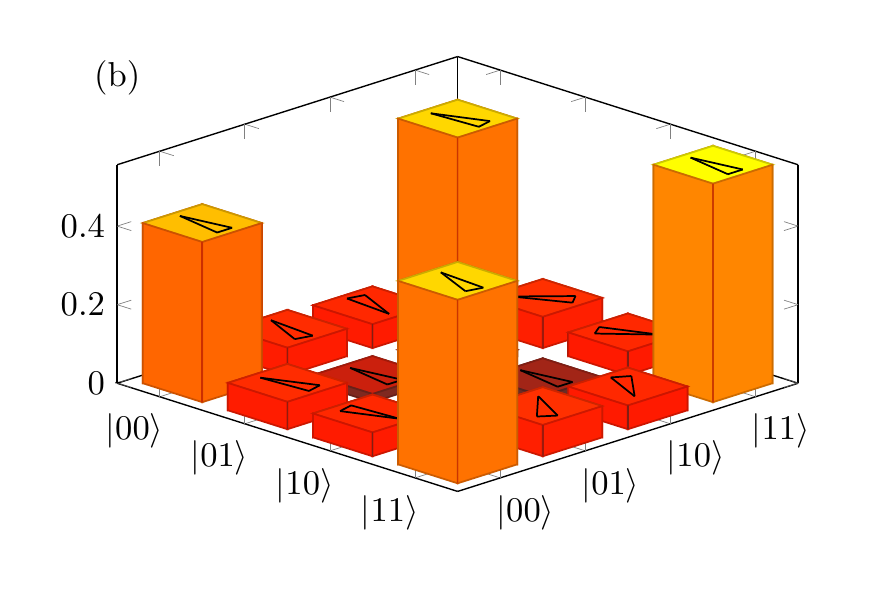} \\
\includegraphics[scale=0.88]{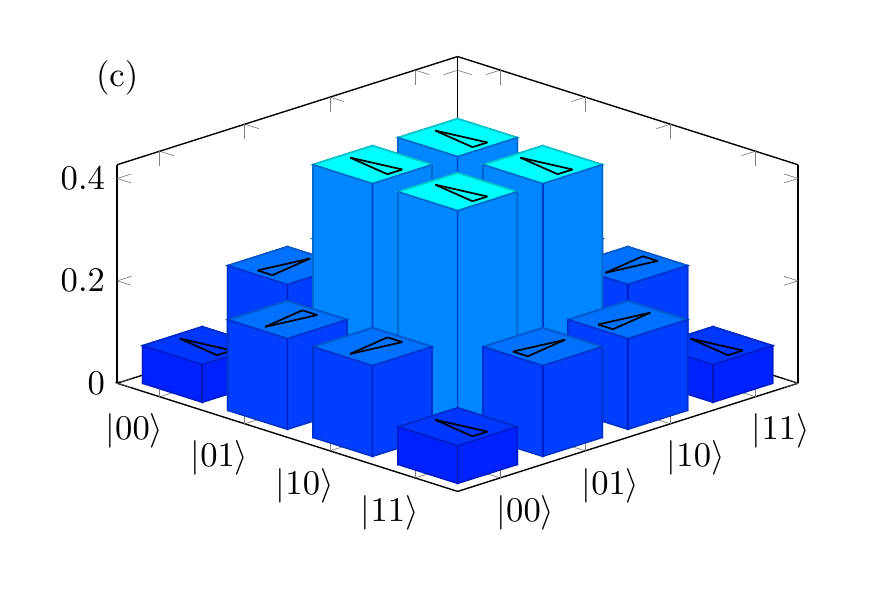} 
\includegraphics[scale=0.88]{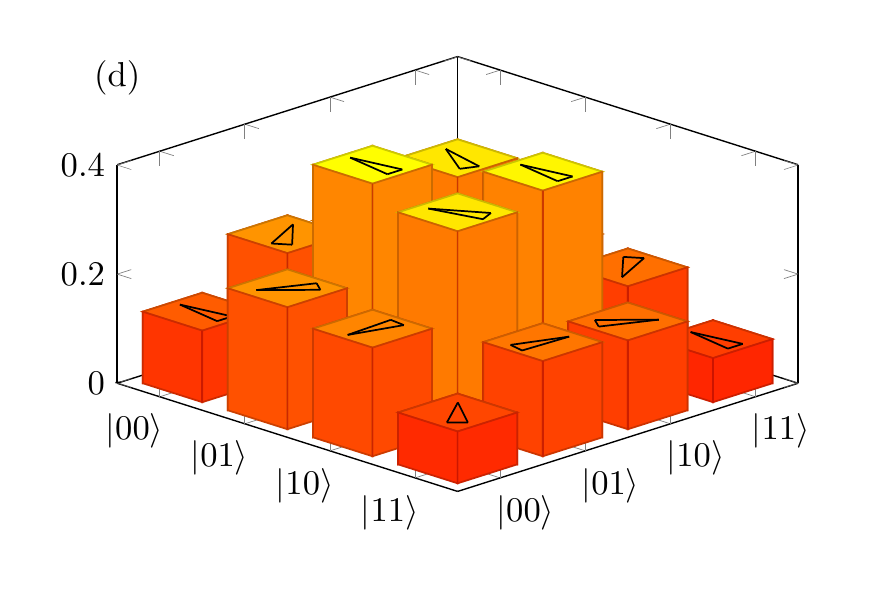}
\caption{\label{fig:supp_res_cuD3pi4} 
(color online) Estimated process matrices of controlled-unitary operation with 
$\varphi=\frac{3\pi}{4}$, $\theta=\frac{\pi}{2}$ and $\alpha=0$: (a) theoretical 
prediction with control qubit set to $|0\rangle$, (b) experimental implementation with 
control qubit set to $|0\rangle$, (c) theoretical prediction with control qubit set to 
$|1\rangle$, (d) experimental implementation with control qubit set to $|1\rangle$. Bar 
height represents the modulus of the matrix element and the black arrow on it's top 
visualize phase shift of that particular matrix element.} 
\end{figure*}

It is fair to derive also inverse relations between both sets of parameters.
By directly comparing the real part of $\chi$ with the real part of the corresponding 
term in $U$, we obtain
\begin{equation}\label{eq:inverse1}
  \cos\frac{\varphi}{2} = \cos\frac{\gamma+\delta}{2}\cos\frac{\omega}{2}.
\end{equation}
Comparing the imaginary part of $\chi$ with the imaginary part of corresponding term
in $U$ and using (\ref{eq:inverse1}) one finds the second inverse relation
\begin{equation}\label{eq:inverse2}
  \cos\theta = 
  \frac{\sin\frac{\gamma+\delta}{2}\cos\frac{\omega}{2}}
  {\sqrt{1-\cos^2\frac{\gamma+\delta}{2}\cos^2\frac{\omega}{2}}}.
\end{equation}
Finally, from (\ref{eq:univ_phase2}) it is trivial to derive the last
inverse relation
\begin{equation}\label{eq:inverse3}
  \alpha = \frac{\delta-\gamma+\pi}{2}.
\end{equation}


\subsection*{Experimental implementation and results}

In our experiment we encode qubits into polarization state where $|0\rangle$ 
corresponds to horizontally polarized $|H\rangle$ and similarly $|1\rangle$ 
corresponds to vertical polarization $|V\rangle$.
We have constructed the experimental setup as shown in Fig. 2 in the main text. 
This setup consists of a tunable c-phase gate enveloped in the signal mode by 
single qubit transformations used to set required unconditional rotations $Z(\pm\alpha)$ 
and $Y(\pm\theta)$. These rotations are implemented by sets of one half-wave plate and 
one quarter-wave plate inserted in front and behind the c-phase gate.
The control state preparation is achieved by individual half-wave plate in the control 
mode since only logical states $|1\rangle$ and $|0\rangle$ are required to switch the 
unitary operation on the signal qubit on and off.

In order to simplify the experimental scheme and to limit the number of required wave plates,
we used only one set of half and quarter-wave plates in front of 
the c-phase gate to set the rotations $Z(\alpha)$ and $Y(\theta)$ and in the same time 
to adjust required input state. Similarly, the half and quarter-wave plates behind the 
c-phase gate are used to implement both the rotations $Z(-\alpha)$ and $Y(-\theta)$ and 
to accomplish polarization projection required for the complete process tomography of
the signal mode.


We tested the operation of our device for six different settings of the parameters 
$\lbrace\alpha,\theta,\varphi\rbrace$. The results of our experiments were summarized 
in Tab.~1 in the main text, where we listed estimated fidelities and purities, that 
are typically about 90\%. Due to the limited space of the main text we plot
examples of the experimental estimated and theoretical matrices of the 
controlled-unitary operation in this section in Fig.~\ref{fig:supp_res_cuD3pi4}. 
The plotted diagrams indicate very good agreement of our experiment with the theoretical 
predictions.

\end{document}